\documentclass[twocolumn,showpacs,preprintnumbers,amsmath,amssymb]{revtex4}
\usepackage{amssymb}
\usepackage[dvips]{graphicx}
\begin{document}

\author{V.V. Kabanov$^{1}$, R.F. Mamin$^{1,2}$, and T.S. Shaposhnikova$^{2}$}
\date{\today}

\affiliation{$^{1}$Jozef Stefan Institute, Jamova 39, 1000
Ljubljana, Slovenia} \affiliation{$^{2}$E.K. Zavoisky Physical
Technical Institute, Russian Academy of Sciences, Kazan, Russia}

\begin{abstract}
Localized charged  states and phase segregation are described in
the framework of the phenomenological Ginzburg-Landau theory of
phase transitions. The Coulomb interactions determines the charge
distribution and the characteristic length of the phase separated
states. The phase separation with charge segregation becomes
possible because of the large dielectric constant and the small
density of extra charge in the range of charge localization. The
phase diagram is calculated and the energy gain of the phase
separated state is estimated. The role of the Coulomb interaction
is elucidated.
\end{abstract}

\title{Localized charged states and phase separation near second order phase transition}
\maketitle

There is a common belief that complex interactions between charge,
spin, orbital, and lattice degrees of freedom may lead to
inhomogeneous ground state with charge and phase separation. The
role of these inhomogeneous states in the anomalous transport and
magnetic properties in
manganites\cite{jin,tokura,millis,nagaev,dagotto,kagan} and in
high temperature superconductors
\cite{gorkov,emery1,egami1,grisha} is often discussed in
literature. It was shown that inhomogeneous states may appear
above Curie temperature and new characteristic temperature $T^{*}$
was introduced\cite{dagotto,gorkov}. Note, that the tendency to
the phase separation is widely discussed for
manganites\cite{nagaev,dagotto,kagan}, as well as for
high-temperature superconductors \cite{gorkov}, microscopic origin
of this anomalous behavior is far from understanding. The origin
and the temperature range of this phenomena is still an open
problem\cite{millis,nagaev,dagotto,kagan,emery1,egami1,grisha,mamin}.
In this paper we discuss the tendency and conditions of the
formation of inhomogeneous states with the spacial charge
localization and the phase separation within the phenomenological
Ginzburg-Landau formalism and clarify the role of the Coulomb
interaction in this phenomena.

The problem of spatially inhomogeneous states in the charged
systems frustrated by the Coulomb interactions is constantly under
debate for different systems
\cite{millis,nagaev,dagotto,kagan,gorkov,muller,emery2,emery3,spivak,muratov,jamei,mertelj}.
Usually the first order phase transition is considered where the
charge density is coupled linearly to the order parameter (as an
external field)\cite{jamei} or to the square of the order
parameter (local temperature)\cite{mertelj}. Very often the
Coulomb interaction is discussed on the qualitative level or on
the last stage of the calculations
\cite{kagan,emery2,emery3,fine}. The general method of
consideration of the long-range Coulomb interaction for the two
dimensional case was proposed in ref.\cite{jamei,mertelj}. In the
recent paper\cite{fine} the effect of the long-range Coulomb
interaction to the second order phase transition was considered.
However the Coulomb interaction was considered only for the
special cases, where the density distribution was approximated in
the attempts to minimize Coulomb contribution to the free
energy\cite{fine}, rather then it was obtained from the general
formulation of the problem. In our opinion, The contribution of
the Coulomb interaction is most important in the determination of
the form of the inhomogeneous states and therefore we consider
this interaction exactly.

Phenomenological approach to the theory of phase transition,
proposed by V.L. Ginzburg and L.D. Landau indicates that the
properties of the system near the phase transition are determined
by the closeness to the phase transition point and weakly
dependent on the other properties of the system. Therefore our
results demonstrates that the properties of the spatially
inhomogeneous states are determined by the closeness to the phase
transition point, but not by the properties of the interactions in
the system. We obtain the contribution of the Coulomb interaction
to the thermodynamic potential of the inhomogeneous state and show
that this contribution may be relatively small, to allow the phase
separation. Note, that the phase separation is possible for the
system displaying phase transition of the second order.

Let us consider doped system in the vicinity of the second order
phase transition. We assume that the average concentration of the
free carriers $\bar{\rho}$ is proportional to the dopant
concentration $\bar{x}$. The thermodynamic potential
$\Phi=\int\phi(\eta,\rho)dV$ describes the behavior of the order
parameter $\eta$ near second order phase transition and
interaction of the order parameter with the charge density.Close
to the phase transition it has the form:
\begin{eqnarray}
\phi(\eta,\rho) &=& \phi_{0}+\phi_{\eta}+\phi_{int}+\phi_{coul}
\nonumber\\\phi_{\eta}(\eta) &=& \frac{\alpha}{2}\eta^{2}+
\frac{\beta}{4}\eta^{4}+ \frac{D}{2}(\nabla\eta)^{2}
\\\phi_{int}(\eta,\rho)&=&\frac{\sigma_{1}}{2}\eta^{2}\rho+
\frac{\sigma_{2}}{2}\eta^{2}\rho^{2} \nonumber \\
\phi_{coul}(\rho)&=&\frac{\gamma}{2}(\rho(r)-\bar{\rho}) \int
\frac{\rho(r^{'})-\bar{\rho}}{|r-r^{'}|}dV^{'}\nonumber
\end{eqnarray}
where $\phi_{0}$ is the density of the thermodynamic potential in
the high-temperature phase, $\phi_{\eta}$ is the density of the
thermodynamic potential in the low-temperature phase, $\alpha$,
$\beta$, $D$ are coefficients in the expansion of the
thermodynamic potentials in powers of the order parameter
($\alpha=\alpha^{'}(T-T_{c})$, where $T_{c}$ critical temperature
in the absence of doping, $\alpha^{'}=1/C$, where $C$ is the Curie
constant), $D$ is proportional to the diffusion coefficient.
Density of the thermodynamic potential $\phi_{int}$ describes the
interaction of the order parameter with the charge, $\sigma_{1}$
and $\sigma_{2}$ are the constants of interactions (the term
$\sigma_{2}\eta^{2}\rho^{2}$ is important to provide global
stability of the system\cite{emery3}, $\sigma_{2}>0$). The density
of the thermodynamic potential $\phi_{coul}$ describes charging
effects due to long-range Coulomb forces, $\bar{\rho}$ is the
average concentration of the charge. We assume that Coulomb
contribution is the strongest. The effect of average dopant
concentration $\bar{\rho}$ to the thermodynamic potential in the
high-temperature phase is included in $\phi_{0}$. The coefficient
$\gamma$ is inversely proportional to static dielectric constant
$\varepsilon$, $\gamma=1/\varepsilon$. As a result the
coefficients in the thermodynamic potential depend on charge
density $\rho$ and the term
$\sigma_{1}\rho(1+\rho\sigma_{2}/\sigma_{1})$ determines the shift
of the critical temperature of the phase transition due to local
charge density. Therefore the influence of the electronic system
on the phase transition manifests itself as a continues shift of
the critical temperature $T_{c\rho}$ with the change of the local
charge density:
\begin{equation}
T_{c\rho}(\rho)=T_{c}-(\sigma_{1}\rho (1+\sigma_{2}\rho
/\sigma_{1}))/\alpha^{'}
\end{equation}
The local charge density may be different from $\bar{\rho}$ in
some region of a finite size.

In the future we perform analysis for 3D case. The analysis for 2D
case is similar. Equilibrium state of the system is determined by
two equations: $\partial\phi(\eta,\rho)/\partial\eta=0$,
$\partial\phi(\eta,\rho)/\partial\rho=0$. The second equation
determines the dependence of the charge density $\rho_{0}$ on
$\eta$:
\begin{eqnarray}
\rho_{0}(\eta) &\simeq& \bar{\rho}+ \frac{\sigma_{1}}{8\pi \gamma}
(1+2\sigma_{2}\bar{\rho}/\sigma_{1})\nabla^{2}\eta^{2}
\nonumber\\&-& \frac{\sigma_{1}^{2}}{32\pi^{2}\gamma^{2}}
(1+2\sigma_{2}\bar{\rho}/\sigma_{1})[ (\nabla^{2}\eta^{2})^{2} +
\eta^{2}\nabla^{4}\eta^{2} ]
\end{eqnarray}
Substituting Eq.(3) to Eq.(1) we obtain the expression of the
thermodynamic potential (1) $\phi_{\rho 0}$ for the equilibrium
distribution of the charge density $\rho_{0}(\eta)$:
\begin{eqnarray}
&&\phi_{\rho 0}(\eta) = \phi_{\eta}(\eta)+ (\sigma_{1}+
\sigma_{2}\bar{\rho})\eta^{2}\bar{\rho}/2 \nonumber\\&-&
\frac{(\sigma_{1}+ 2\sigma_{2}\bar{\rho})^{2}} {32\pi \gamma} [
(\nabla \eta^{2})^{2} -\frac{\sigma_{2}}{4\pi \gamma}
\eta^{2}(\nabla^{2}\eta^{2})^{2} ]
\end{eqnarray}
The negative sign in the third term of Eq.(4)
($-(\nabla\eta^{2})^{2}$) indicates that uniform state may be
unstable towards inhomogeneous fluctuations. Similar instabilities
were discussed in the case of first order phase transition in
Refs.\cite{jamei,mertelj}.

\begin{figure}
\begin{center}
\includegraphics[angle=-90,width=0.45\textwidth]{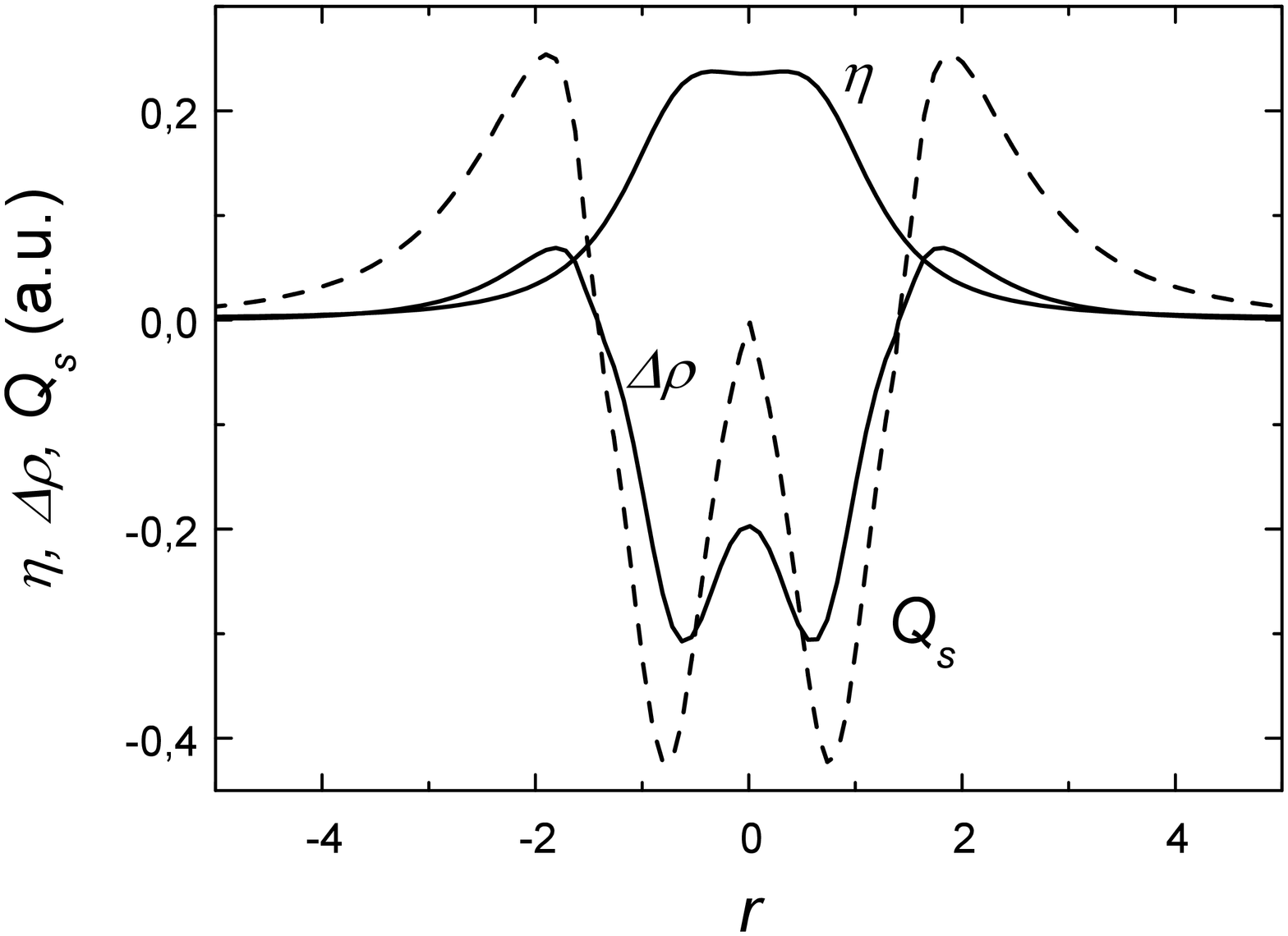}
\vskip -0.5mm \caption{(a) Spatial variations of the order
parameter $\eta(r)$ and charge density
$\Delta\rho(r)=\rho(r)-\bar{\rho}$ along the diagonal of the
sphere in the equilibrium nonhomogeneous state, obtained by
numerical minimization of the free energy; $Q_{s}$ is integral
concentration of the charge as a function of distance from the
center of sphere ($Q_{s}=\Delta\rho(r)S_{r}$)}
\end{center}
\end{figure}

This instability leads to spatially inhomogeneous solutions. Most
simple solution is spherically symmetric. This equilibrium
inhomogeneous state, obtained by computer minimization of the
thermodynamic potential Eq.(1), has spherical form and plotted in
Fig.(1). This results are obtained by applying the numerical
technique described in Ref.\cite{miranda}. This solution has
characteristic size $R_{0}$ and characteristic charge density
inside of this region $\rho_{0}$ under condition that average
charge density in the system is equal to $\bar{\rho}$. It is clear
that the charge is concentrated near the surface of the sphere. It
allows to use approximation of the double electrical layer for
calculation of the Coulomb energy. At large distances from the
sphere the order parameter and the charge density are equal to
their equilibrium values $\rho=\bar{\rho}$ and
$\eta=\eta_{s}(\bar{\rho})$ ($\eta_{s}=0$ or
$\eta_{s}=-\tilde{\alpha}(\bar{\rho},T)/\beta$). As a result the
density of the thermodynamic potential $\phi_{s}$ for this
solution we can write in the form:
\begin{equation}
\phi_{s}(R_{0},x_{0})=\phi_{0}-{A(x_{0})\over{3}}R_{0}^{3}+
{B(x_{0})\over{2}}R_{0}^{2}+ {K(x_{0})\over{4}}R_{0}^{4}
\end{equation}
Here we use dimensionless concentration $x$ instead of charge
density $\rho$ ($x=\rho a^{3}/e$, $a$ is the lattice constant, and
$e$ is elementary charge) and $\bar{x}$ corresponds to the level
of doping. The coefficients in this formula are defined as
follows: $A(x_{0})=\pi(\tilde{\alpha}(x_{0},T))^{2}/\beta V_{0}$,
$B(x_{0})\simeq 8\pi D (-\tilde{\alpha}(x_{0},T))/d\beta V_{0}$,
$K(x_{0})\simeq \gamma e^{2}dF(\bar{x}-x_{0})^{2}/a^{6}V_{0}$
($\tilde{\alpha}(x_{0},T)=\alpha(T)+
\tilde{\sigma_{1}}x+\tilde{\sigma_{1}}x^{2}$,
$\tilde{\sigma_{1}}=e\sigma_{1}/a^{3}$,
$\tilde{\sigma_{2}}=e^{2}\sigma_{2}/a^{6}$, $F$ is dimensionless
factor which determines charge distribution
$F=4a^{6}\int\phi_{coul}dV/\gamma e^{2}
R_{0}^{4}d(\bar{x}-x_{0})^{2}$, $F=1/18$ in the limit of the
double electrical layer, $V_{0}$ is the elementary layer that
contains one sphere ($V_{0}=V/n$ where $n$ is the number of
spheres), d is the thickness of the interphase boundary. Note,
that Eq.(5) is valid if characteristic dimension of the
nano-regions is larger then characteristic length $\xi=\hbar
v/k_{B}T_{c\rho}(x_{0})$, where $\hbar$ is the Plank's constant,
$k_{B}$ is the Boltsmann constant, $v$ is the characteristic
velocity of the problem (in the case of structural phase
transitions it is equal to the sound velocity and in the case of
electronic phase transitions it is equal the Fermi velocity). This
may be achieved relatively easy if $T_{c\rho}(x_{0}) > 10$K.  Here
we take into account the strong screening of the localized charges
and write the energy in the double electrical layer approximation.
In that case it is proportional to $dR_{0}^{4}$. If the screening
is absent this energy is proportional to $R_{0}^{5}$\cite{gorkov}.
The parameters $A(x_{0})$ and $B(x_{0})$ depend on temperature
$T$, and parameter $K(x_{0})$ depends on the average charge
density $\bar{x}$. Parameters $(\bar{x},T)$ are external and we
find the phase diagram in the space of these parameters. Since
$\bar{x}$ is determined by the level of doping, we study the
change of the properties of the system with doping.

The conditions for the minimum of the potential
$\phi_{s}(R_{0},x_{0})$ as a function of $R_{0}$ and $x_{0}$ and
the condition that minimum of the potential is lower then the
potential of the uniform state $\phi_{s}<\phi_{0}$ define the
following set of equations:
\begin{eqnarray}
\partial\phi_{s}(R_{0},x_{0})/\partial R_{0}=0 \\
\partial\phi_{s}(R_{0},x_{0})/\partial x_{0}=0 \\
\phi_{s}=\phi_{0}
\end{eqnarray}
These equations define the upper boundary of the existence of the
inhomogeneous state. Eq.(6) defines the equilibrium size of the
charged domain $R_{0}$:
\begin{equation}
R_{0}={A(x_{0})\over{2K(x_{0})}}(1+\sqrt{1-
{4K(x_{0})B(x_{0})\over{A(x_{0})^{2}}}}
\end{equation}
On the upper boundary of the region of the inhomogeneous states
(Eq.(8)) we obtain the localization radius
$R_{0}=2A(x_{0})/3K(x_{0})$. As a result Eq.(8) gives the
following equation for $x_{0}$ as a function of temperature $T$:
\begin{equation}
\tilde{\alpha}(x_{0},T)(\sigma_{1}+2\sigma_{2}x_{0})^{2}+
{16\gamma e^{2}\beta F D\over{\pi a^{6}}}=0
\end{equation}
Using Eqs.(10) and (7) we write equation for upper boundary of the
inhomogeneous states in the space of external parameters
$(\bar{x},T)$ (the sign $\pm$ is determined by the sign of
$\sigma_{1}$):
\begin{equation}
\bar{x}=x_{0}(T)\pm ({\pi a^{6}\over{36\gamma e^{2}\beta F
D}})^{1/2}\tilde{\alpha}^{3/2}(x_{0}(T),T)=0
\end{equation}

\begin{figure}
\begin{center}
\includegraphics[angle=-90,width=0.45\textwidth]{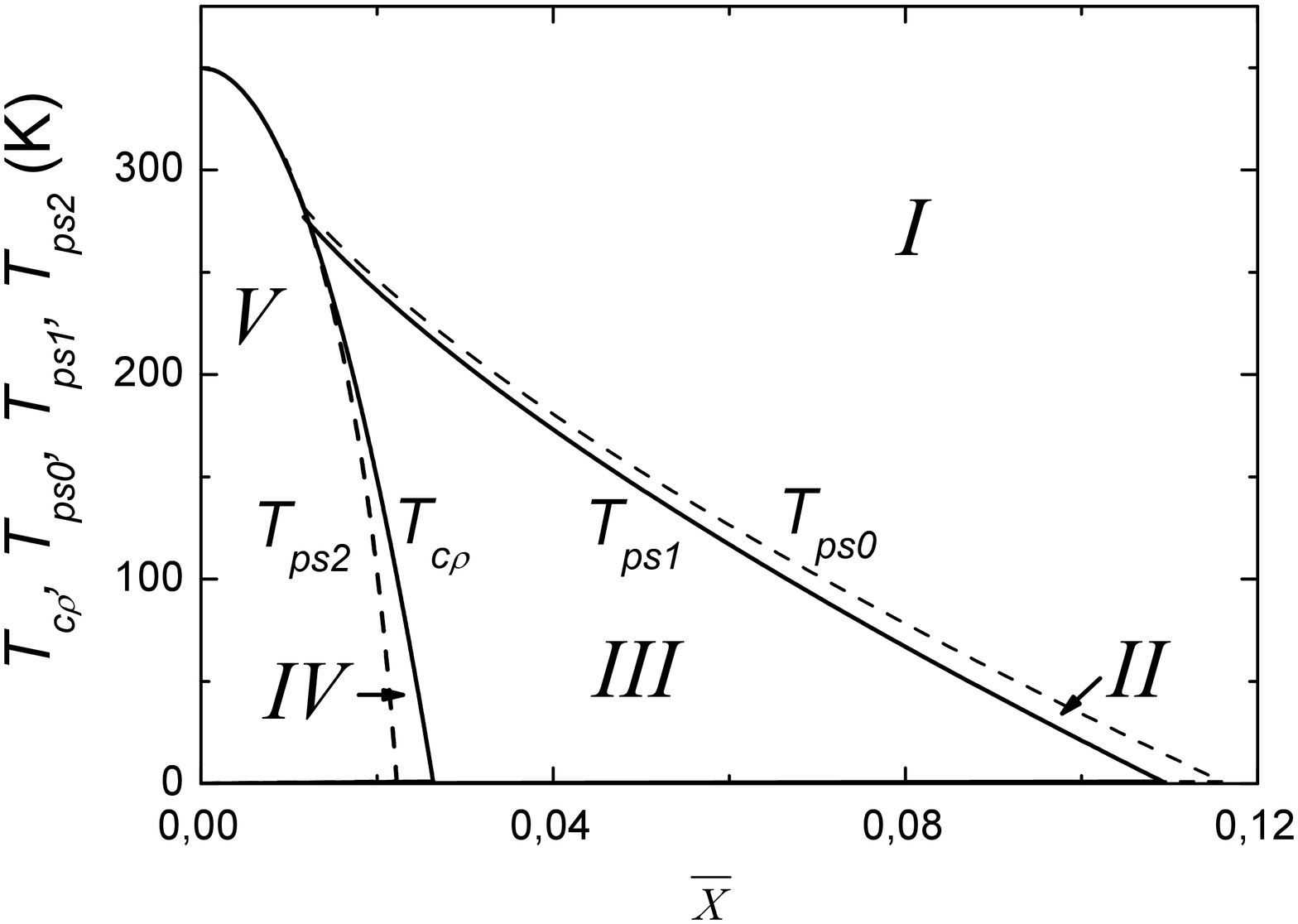}
\vskip -0.5mm \caption{Phase diagram of the system with
$T_{c}=350K$, $\sigma_{1}/\alpha^{'}=50K$,
$\sigma_{2}/\alpha^{'}=5 10^{5}K$, $\gamma e^{2}\beta FD/\pi
a^{6}\alpha^{'}=2.33T_{c}^{3}$: $I$ is is the region of the stable
high-temperature phase; $II$ is the region of the stable
high-temperature phase with metastable bubbles of the
low-temperature phase; $III$ is the region of the stable
inhomogeneous phase and metastable homogeneous high-temperature
phase; $IV$ is the region of the stable inhomogeneous phase and
metastable uniform low-temperature phase; $V$ is the region of the
stable uniform low-temperature phase.}
\end{center}
\end{figure}

These equations define the temperature $T_{ps1}$ of the transition
to the region of stable inhomogeneous states in the space of
external parameters $(\bar{x},T)$ ($T_{ps1}=f_{1}(\bar{x})$) where
the energy of this inhomogeneous state is lower then the energy of
homogeneous state. Similar consideration leads to the equation
which defines lower boundary of the inhomogeneous states
$T_{ps2}=f_{2}(\bar{x})$. Note that lower boundary $T_{ps2}$ will
be always close to the temperature $T_{c\rho}(\bar{x})$, because
we do not gain, but contrary loos in energy $\phi_{\eta}$ when the
bubble of the high-temperature phase is formed
($\phi_{\eta}(\eta=0)=0$, and $\phi_{\eta}(\eta_{0})<0$). It leads
to the considerable difference between the formation of the bubble
with the low-temperature phase in the high-temperature region and
the bubble of the high-temperature phase in the low-temperature
region. The first one is energetically favorable and therefore the
region of the existence of this bubbles is considerably larger. It
is important to note that Eqs.(10,11) which determine the phase
diagram do not depend on $d$. This fact allows us to avoid
optimization of our analysis with respect to $d$. Typical phase
diagram of the inhomogeneous state is presented in Fig.2 for the
case of of positive $\sigma_{1}>0$. Phase transition to the
nonhomogeneous state represents typical phase transition of the
first order. Metastable inhomogeneous phase appears at the
temperature $T_{ps0}$ which is much higher then the temperature of
phase transition and it is shown in the phase diagram by dashed
line. This line is determined from the conditions Eqs.(6,7) and
$A(x_{0})^{2}=4K(x_{0},\bar{x})B(x_{0},\bar{x})$ instead of
Eq.(8). The characteristic size of the charged regions in that
case is determined by the condition
$R_{0}=A(x_{0})/2K(x_{0},\bar{x})$.

Very often the Coulomb energy is estimated as $e^{2}/a\simeq4eV$,
which corresponds to two elementary charges localized in
neighboring unite cells. However in our case to make the phase
separation possible it is sufficient to have only small variation
of the charge density per unite cell in comparison with average
charge density $a^{3}(\rho_{0}-\bar{\rho})\simeq0.1-0.2e$.
Therefore we can estimate the Coulomb contribution to the free
energy $u_{\rho}$ as well as the energy gain due to formation of
the low-temperature phase $u_{\eta}$ (the second and the forth
terms in Eq.(5)) per one unite cell in the nonhomogeneous phase
$u_{i}=3|\phi_{i}|a^{3}V_{0}/4\pi R_{0}^{3}$. We assume that $d$
is small and neglect surface term in Eq.(5). For the Coulomb
contribution we obtain:
\begin{equation}
u_{\rho}=\Bigl({3e^{2}(x_{0}-\bar{x})^{2}\over{4\pi \varepsilon
a}}\Bigr) {dR_{0}F\over{a^{2}}}
\end{equation}
The characteristic energy in the brackets is of the order
$u_{0}=0.0003-0.001eV$, here we use static dielectric constant
$\varepsilon=30-40$, which is consistent with the measurements on
manganites and high-temperature superconductors. The increase of
the bubble size leads to the increase of the energy $u_{\rho}$ by
the order of magnitude or more in comparison with $u_{0}$, but on
the other hand geometrical factor $F$ decreases $u_{\rho}$.
($F=1/18$ in the limiting case of the double electrical layer).
Therefore $u_{\rho}$ is evaluated as $0.003-0.01eV$. The
corresponding estimate for the energy gain due to formation of the
low-temperature phase is $u_{\eta}\simeq 3k_{B}T_c/4$
($u_{\eta}\simeq a^{3}(\tilde{\alpha}
(x_{0},T=0))^{2}/4\beta\simeq
a^{3}T_{c}\eta_{0}^{2}(x_{0},T=0)/C$, where $C=N(\eta_{0}(T=0)
a^3)^{2}/3 k_{B}$ and $N=a^{-3}$. Therefore $u_{\eta}\simeq
0.02-0.05eV$ for $T_{c}\simeq 270-700 K$. Therefore in our case
the phase separation becomes possible. The typical length scale of
the nano-regions is estimated as $R_{0}\simeq 1-20nm$.

The analysis of the pair distribution function obtained by neutron
scattering shows that the charge density in manganites is
localized on the scale of 3 to 4 interatomic distances.  This is
the case even if holes are localized in the form of
polarons\cite{dagotto,louca}. It confirms that holes are localized
not in the form of well separated polarons but in the form of
nanoregions with finite charge density of polarons. The extra
charge in that case is not more then $0.1-0.2e$ per unite cell.
This is in consistence with our estimates. This state is
characterized as the state with nano-dimensional charge and phase
separation. Dynamics of these charged nano-regions may lead to
high value of the dielectric constant in the low frequency range
\cite{mamin}.

Therefore we have shown that the second order phase transition
with critical temperature equal to zero at the certain doping
(quantum critical point) is unstable with respect to formation of
the spatially inhomogeneous charged states. Within the
phenomenological Landau theory  we have shown that these states
appears at some temperature $T_{ps1}$, which is substantially
higher then the temperature $T_{c\rho}$ (Fig.2). As a result the
phase transition becomes effectively first order phase transition.
Note that the Coulomb interaction determines the charge
distribution, the screening and the characteristic length scale of
the nonhomogeneous states. The spatially inhomogeneous state
becomes possible in the systems with the large dielectric
constants (30-40) and with relatively small charge density
variations. The nonhomogeneous states were discussed also in the
case of the phase transitions of the first
order\cite{jamei,mertelj}. However in that case spatial variations
are related to the existence of two minima in the free energy in
the region of hysteresis of the first order phase transition. In
our case the nonhomogeneous states are related to the spatial
charge redistribution.

In conclusion we note that the localized charged states and the
phase separation appears even in the case of the second order
phase transition. Properties of these states are described within
the phenomenological theory of the phase transitions. The Coulomb
interaction determines the spatial charge distribution, the
screening and the characteristic length of charge localization.
The charge and the phase separation becomes possible because of
large dielectric constants and relatively small spatial variation
of the charge density. We derive phase diagram and estimate
different contributions to the free energy.

Enlightening discussions with A. P. Levaniuk, B.Z. Malkin,  and D.
Mihailovic are highly appreciated. We acknowledge financial
support from Russian Foundation for Basic Research (Grants \#
05-02-17182 and \# 08-02-00045), Slovenian Ministry for Science
and Technology and Ad-Futura (Slovenia).

\end{document}